\documentstyle[revtex]{aps}

\begin{document}
\draft
\begin{center}
\Large\bf

Inverses of slopes of invariant inclusive spectra
of emitted protons and $\pi^-$-mesons in ${}^{4}HeC$
and ${}^{12}CC$ interactions with the total disintegration of nuclei.

\vspace{1.0cm}

\large

M. K. Suleimanov\cite{mais} , O. B. Abdinov

\large\it  Physics Institute, Azerbaijan Academy of
Sciences, Baku, Azerbaijan Republic

\vspace{1.0cm}

\large
A.I.Anoshin

\large\it
Nuclear Physics Institute, Moscow State University, Moscow, Russia

\vspace{1.0cm}

\large
J.Bogdanowicz

\large\it
Soltan Institute for Nuclear Studies, Warzaw, Poland

and Joint Institute for Nuclear Research, Dubna,Russia
\vspace{1.0cm}

\large
A.A.Kuznetsov

\large\it
Joint Institute for Nuclear Research, Dubna,Russia

\vspace{1.0cm}
\end{center}

\newpage

{\large

 The ideas  that extreme states of nuclear matter arise in
events with total disintegration of nuclei (TDN) and
as these states arise, the properties of events qualitatively change
with the number of protons emitted from the nucleus ($Q$), starting from
its certain boundary number ($Q^*$),
are used in this paper for the experimental
search for extreme states of nuclear matter. For realization of these
ideas, the invariant inclusive spectra of protons and $\pi^-$-mesons as
a function of their kinetic energies $T$ in the lab system for
${}^4HeC$ and ${}^{12}CC$ interactions at the momentum 4.2 A GeV/c with
different values of $Q$ are used. The spectra are fitted by the
expressions of the form $\sum_{i=1}^{n} a_i\exp(-b_iT)$ and the $Q$-
dependencies of the inverses of slopes $T_i=1/b_i$ are
studied.  It is found that these spectra have two components (the
low energy component  corresponds to $T_1$ and the high energy component
corresponds to $T_2$ ) and contain the regime change points at $Q\ge $
2 and 4 for ${}^4HeC$ interactions and at $Q\ge$ 6 and 9 for ${}^{12}CC$
interactions .  For protons the first component is mainly connected with
the evaporation protons and the leading-stripping fragments produce
great influence on $T_2$. In the TDN region the  leading- stripping
effect is suppressed and the values of $T_i$ for
$\pi^-$-mesons begin to increase with  increasing  number of
protons. We consider this increase to be a  signal from the extreme
states of nuclear matter.  The value of the "temperature" of these
states is about 0.140 GeV.

\noindent The work has been performed at the Laboratory of High
Energies, JINR.}

\vspace{0.5cm}

{\large\noindent PACS number(s): 25.70.Np, 25.45.-z, 22.55.-e
}

\begin{center}
{\Large\bf I. INTRODUCTION}
\end{center}

{\large

 In this paper we discuss the experimental results of studying the invariant inclusive
spectra $f=(…/\sigma) d^{3}\sigma/dp^{3}$ of protons and $\pi^-$-mesons
as a function of their kinetic energies $T$ in the laboratory system of coordinates.
The spectra $f(T)$ were obtained for ${}^4HeC$ and ${}^{12}CC$ interactions
at the momentum 4.2 GeV/c with a different number of protons.

The aim of this paper is to reveal experimentally the extreme states of nuclear
matter in  interactions of relativistic light nuclei. The investigation  of
the states is the most important trend in relativistic nuclear physics. The
primary aim of these investigations is  experimental observation of the
theoretically predicted  phenomenon of nuclear matter deconfinement and
transition to the  quark-gluon plasma [1].

It is  assumed that these states can arise at definite (critical)
values of energy density $\varepsilon$ (~$\varepsilon_c$~) or "temperature"
$T_0$ (~$T_{0c}$~) . The values of $T_0$ can be  determined
only in definite kinematic regions ( e. g., in the region of large angles
or large $p_t$ , etc.) and in the framework of a some model-dependent assumptions
(e. g., of the chemical potential etc.). Experimental determination
of $\varepsilon$  is practically impossibile (because of the absence
of information on the size of the radiation region). It can only be
calculated in the framework some theoretical models. The number of participant
nucleons of colliding nuclei is the experimentally determinable quantity
closest to  $\varepsilon$ [3]. This quantity also cannot be
unambiguously  found in the  experiment because of the absence of full
information on the evaporation protons  and stripping fragments.

We used the following ideas for the experimental search for the extreme states of
nuclear matter :

1. These states can arise in events with the maximum number of protons  $Q$ and
therefore they are accompanied  by the processes of total disintegration
of colliding nuclei (TDN).

2. As  these extreme states arise, the properties of events must qualitatively change with $Q$
beginning with a definite boundary value $Q \rightarrow Q^*$ (the experimentally obtained values of $Q^*$ will
characterize the values of $\varepsilon_c$).The experimental
results obtained in [4] point to validity of this idea and the
values of $Q^*$ were used for selection of events with TDN on condition
that $Q\ge Q^*$.

Experimentally, these ideas are realized by  studying the dependence of
different characteristics of nucleus-nucleus interactions on $Q$.

In the present paper, to search for the extreme states of nuclear matter the
$Q$-dependence of $T_i$ - the inverses of the slopes of $f(T)$ -  were studied.
Among the experimentally determinable variables  $T_i$ is the closest to
$T_0$. This quantity can be represented as a sum of energies
corresponding to $T_0$ and the energy due to motion of the system
itself. One can expect that in the region of large $Q$ the motion energy
will tend to zero. Therefore, on transition to the studied
region of large $Q$ - to
processes of TDN - the values of $T_i$ will become close to $T_0$.

For this, the spectra of $f(T)$ were separately studyied for events with
different numbers of protons $Q$. It is known that ( see for example [5])
the spectra $f(T)$  have the exponential form with a good accuracy .
Therefore, to study the dependence of $f(T)$ on  $Q$ we fitted these spectra
by the expressions of the form

\begin{equation}
\sum_{i=1}^{n} a_i\exp(-b_iT).
\end{equation}

(here $a_i$ and $b_i$ are the fitting parameters  . The quantity $n$ is
defined from the condition of the best fitting - the errors in  defining
$a_i$ and $b_i$ and the values of $\chi^2$  per degree of freedom are
the smallest) and studied the dependence of $T_i=1/b_i$ on the variable $Q$.

\vspace{1.0cm}

\begin{center}
{\Large\bf II. EXPERIMENTAL DETAILS}
\end{center}
\vspace{1.0cm}

{\large

In this paper the experimental data were obtained by exposing the 2-m propane
bubble chamber of  LHE JINR to the beams of
light relativistic nuclei at the Dubna Synchrophasotron at a momentum of
4.2 A GeV/c . The chamber was placed in a magnetic field of 1.5 T.  The
statistics are 4852  ${}^4HeC$ and  7327   ${}^{12}CC$
interactions ( for methodical details see [6]) . We had the 4$\pi$
geometry for measurement and detected practically  all secondary particles.
However, it is necessary to note  that in our experiment  protons
are identified by ionization and their range in the   momentum interval
0.15-0.50 GeV/c .  The protons with the momenta $ p < 0.15$ GeV/c have
the range less than 2 mm and most of them are not seen in a photo. For all
positive particles with the momenta higher than 0.5 GeV/c we introduced the weight
that determined the probability for the particle to be a proton or a $\pi^+$-
meson.  The characteristics of the $\pi^+$-meson were used for determination of
the weights. The smallest momentum for detection of $\pi^-$-mesons was 0.07 GeV/c.
The contamination by electrons and by negative strange particles
did not exceed 5$\%$ and 1$\%$, respectively.

In this paper  the variable $Q$ is used to determine the number of protons.
We defined  $Q$ for each event as

\begin{equation}
\ Q=N_+ - N_{\pi^{-}}.
\end{equation}

where $N_+$ is the number of positive particles and $N_{\pi^{-}}$ is the
number of $\pi^{-}$ - mesons  ( we assumed $N_+ = N_{\pi^{-}}$ ).
Experimental losses of particles and
errors in identification of  secondary particles influenced the accuracies
of determination of $Q$.  The poor accuracy in determination of $Q$
can lead to appearance of "wrong" $Q^*$ and to expansion of the regime  change
regions. For this reason we cannot determine the number of the regime change
regions and the exact values of $Q^*$. For decreasing the influence of this
circumstance, we do not consider the groups of events with definite values of
$Q$ but we consider the groups of events with $Q \ge $. It is known that
setting up the experiment like this one decrases the influence of the
accidental processes. Therefore, we divided the available statistical material into  groups of events with the
following values of Q:

\begin{equation}
\ Q\geq 1 ;2 ; 3 ; 4; 5;6; 7.
\end{equation}

for ${}^4HeC$ interactions  and

\begin{equation}
\ Q\geq 1 ;2 ; 3 ; 4; 5; 6; 7; 8; 9; 10; 11.
\end{equation}

for ${}^{12}CC$ interactions and
for each group of events the spectra of $f(T)$ were obtained.

\vspace{1.0cm}

\begin{center}
{\Large\bf III. THE RESULTS AND DISCUSSION}
\end{center}
\vspace{1.0cm}

{\large

The spectra of $f(T)$ for ${}^{12}CC$ with $Q \ge 1;3;6;9$ are shown in figs.
1a,b.(solid lines demonstrate the results of the best fitting ). One can see
that these spectra have an exponential form. The spectra obtained at other
values of $Q\ge$  and the spectra for ${}^4HeC$ interactions have a similar form.
Therefore, all spectra were fitted by  expression (1).

It turned out that $n=2$ in all  cases considered ( in table 1 the
values of $\chi^2$ per degree of freedom are given), i.e., these spectra have
two components - the low-energy component corresponding to $T_1$ and the high-
energy component corresponding to $T_2$. This result is a well known experimental
fact ( see for example [5] and references therein) and is interpreted as an
indication of the presence of two sources ( or mechanisms) of emission of the
observed  particles.

The $Q$-dependence of $T_i$  is shown in figs. 2a,b-3a,b .
It is seen that for protons

 - the values of $T_1$ (fig. 2a) reach  30 MeV at maximum and weakly
decrease with growing  mass of the projectile nucleon -  interaction  energy.
This result agrees with the conclusions  [5] that the
low-energy part of the proton spectra is mainly connected with the
evoparation protons . The decrease in $T_1$  with growing
$Q$ is due to increasing  energy loss  for secondary interactions ;

 - the values of $T_2$ (fig. 2b) sharply increase with growing  mass of the
projectile nucleon - interaction energy .

  The regime change points are observed  at  $Q \ge $ 4 and 6
for ${}^4HeC$ interactions and $Q\ge$ 8 for ${}^{12}CC$ interactions. This
result confirms our
basic idea ( see Introduction) and is in good agreement with the  conclusions
[6], where similar regime change points were found from the analysis of the
probability of $Q$ distributions of events  and the dependencies
of the mean characteristics  of events on $Q$. In [6] the values of
$Q$ corresponding to the regime  change points were regarded as "critical"
($Q^*$) and were  used for separating events with  total disintegration
of nuclei (TDN) for $Q\ge Q^*$.  Note that in the region
$Q\le Q^*$ the values of $T_2$ for protons are almost independent of $Q$ and
they sharpply decrease in the $Q\ge Q^*$ region. We think that this is
connected  with the admixture of leading-stripping fragments to the  protons
considered. In the region of large $Q$  the leading-stripping effect
is suppressed. Against the background of this effect it is difficult to get
information on the inner energy (and thus on $T_0$) of nuclear matter since
the difference between the values of $T_0$ and leading-stripping particle
energy is very large.

From the data in fig. 3a,b it is evident that

- the spectra of secondary $\pi^-$-meson are also well fitted by the
expression with $n=2$ (i.e., there are two sources (mechanisms) of
$\pi^-$-mesons emission) and contain the regime change points for
${}^4HeC$ interactions at  $Q \ge $ 2 and 4 (fig. 3b) and at $Q \ge $ 6
and 9 for ${}^{12}CC$ interactions (see fig. 3a).

- at the beginning of this spectra
of $T_i$ decreases with $Q$ increasing to $Q^*$ and reach its minimum; then, as
$Q$ increases  in the $Q\ge Q^*$ region, $T_i$ stops decreasing for ${}^4HeC$
interactions and sharply increases for ${}^{12}CC$ interactions.
It is interesting to note that $T_1$ and $T_2$ similarly depend on $Q$. This
indicates that the high-energy and low-energy components of $\pi^-$-mesons
differ by that the latter are emitted from the source nearly at rest and
therefore they can characterize the states of nuclear matter. We think that
it is the  values of $T_1$ in the region $Q\ge Q^*$ that
correspond to the "temperature" of nuclear matter.  Its maximum value
$T_1 \simeq 0.140 $ corresponds to the "temperature" of nuclear matter in extreme
states. If $T_i$ does not increase in the region $Q\ge Q^*$ for ${}^4HeC$
interactions, it means that  the necessary density of
energy is not reached in these interactions ;

- a decrease in $T_i$  with increasing $Q$ at the beginning of the spectra is
due to an increase in secondary interactions and rescattering of
$\pi^-$-mesons from intranuclear nucleons, which results in $\pi^-$-meson
energy losses.  In the $Q\ge Q^*$ region the loss of  $\pi^-$-meson
energy becomes  insignificant. This means that in this region the processes of
meson production occur simultaneously with emission of protons, i.e,
the collective processes dominate.

\vspace{1.0cm}

\begin{center}
{\Large\bf ACKNOWLEDGEMENTS}
\end{center}
\vspace{1.0cm}

{\large
    The authors consider it to be their  pleasant duty  to thank the
staff of the 2-m propane bubble chamber  for providing the experimental material
and also to Academician A. M. Baldin for his continuous attention to the work.
}

\vspace{1.0cm}

\begin{center}
\line(1,0){250.}
\end{center}
\vspace{1.0cm}

\begin{itemize}

\item [[1.]] J. W. Harris and STAR Collaboration, in: Proceedings X
International Conference on Ultra-Relativistic Nucleus-Nucleus Collisios,
Borlange, Sweden, June 20-24, 1993; Nucl.Phys., A{\bf 566} , 277 (1994);

J.Schukraft and ALICE Collaboration, in: Proceedings X
International Conference on Ultra-Relativistic Nucleus-Nucleus Collisios,
Borlange, Sweden, June 20-24, 1993; Nucl.Phys., A{\bf 566} , 311, (1994).

\item [[2.]] Sh. Nagamiya {\it et al.}, Phys.Rew. C , {\bf 44} , 971, (1981) ;

B. Brockmann {\it et al.} Phys. Rew. Lett., {\bf 53}, 2012, (1984) ;

S.Backovic, D. Salihagic {\it et al.} Phys.Rew. C, {\ bf 46} , 1501, (1992);

L. Ahle, Y. Akiba {\it et al.} Phys. Rew. C , {\bf 55}, 2604, (1997)

\item [[3.]] C.De Marzo,M. De Palma {\it et al.} Physical Rewiew D, {\bf 29}
, 2476, (1984);

Lj. Simic, S. Backovic {\it et al.} Z. Phys. C,  Particles and Fields, {\bf 48},
577, (1990)

\item [[4]] O. B. Abdinov {\it et al.} , JINR Rapid Communications, No. {\bf 1[75]} , 51 , (1996);
O. B. Abdinov {\it et al.} , JINR Rapid Communications  No. {\bf 1[81]} ,109 , (1997) .

\item [[5.]] A.M. Baldin {\it et al.} Communication of JINR , No. {\bf  P1-11302} ,
Dubna 1978 ;  V. I. Komorov, H. Muller, Communication of JINR ,
No. {\bf E2-12439}, Dubna 1978 .

\item [[6.]] N.Akhababian {\it et al.}  Preprint JINR No. {\bf 1-12114},Dubna,1979.

\end{itemize}

\newpage

\vskip 0.5cm

\figure{$T$-dependence of invariant inclusive spectra for
protons (a) and $\pi^-$-mesons (b) emitted in ${}^{12}CC$ interactions.
Solid lines are the fitting results . The data are given
for events with $Q\ge 1$ ,  $Q\ge 3$ (the corresponding values
of $f(T)$ are divided by 100) $Q\ge 6$ (the corresponding values of
 $f(T)$ are divided by 10000), $Q\ge 9$ (the corresponding values of
 $f(T)$ are divided by 1000000).
\label{fig1}}

\figure{
 $Q$-dependence of $T_1$ (a) and $T_2$ (b)
for protons emitted in ${}^4HeC$ ($\bullet$) and ${}^{12}C$ ($\circ$)
interactions.
\label{fig2}}

\figure{ $Q$-dependence of $T_1$ (a) and $T_2$ (b)
for $\pi^-$-mesons emitted in ${}^4HeC$ ($\bullet$) and ${}^{12}C$
($\circ$) interactions.
\label{fig3}}

\newpage

\begin{center}
{\Large  TABLE I. The values of $\chi^2$ per degree of freedom}
\end{center}

\begin{center}

\line(1,0){125.}
\vspace{0.1cm}

\begin{tabular}{ccccc} \hline
\multicolumn{1}{c}{Q}&\multicolumn{2}{c}{${}^4HeC$}&\multicolumn{2}{c}{${}^{12}CC$}\\ \hline
\multicolumn{1}{c}{} &\multicolumn{2}{c}{$\pi^-~~~~~~p$}&\multicolumn{2}{c}{$\pi^-~~~~~p$}\\ \hline
1 &1.17&2.64&0.61&2.29 \\
  2 & 0.60 &2.45 &0.56 &2.18 \\
  3 & 1.33 &2.10 &0.64 &1.94 \\
  4 & 0.90 &2.33 &0.64 &1.70 \\
  5 & 0.73 &2.01 &0.73 &1.44 \\
  6 & 0.47 &0.93 &0.58 &1.22 \\
  7 &  0.81 &1.02 &0.54 &0.92 \\
  8 &   -    &  -   &0.55 &1.72 \\
  9 &   -    &  -   &0.39 &1.35 \\
 10 &   -    &  -   &0.61 &1.65 \\
 11 &   -    &  -   &0.56 &1.54 \\
\hline
\end{tabular}

\vspace{0.1cm}
\line(1,0){125.}
\end{center}

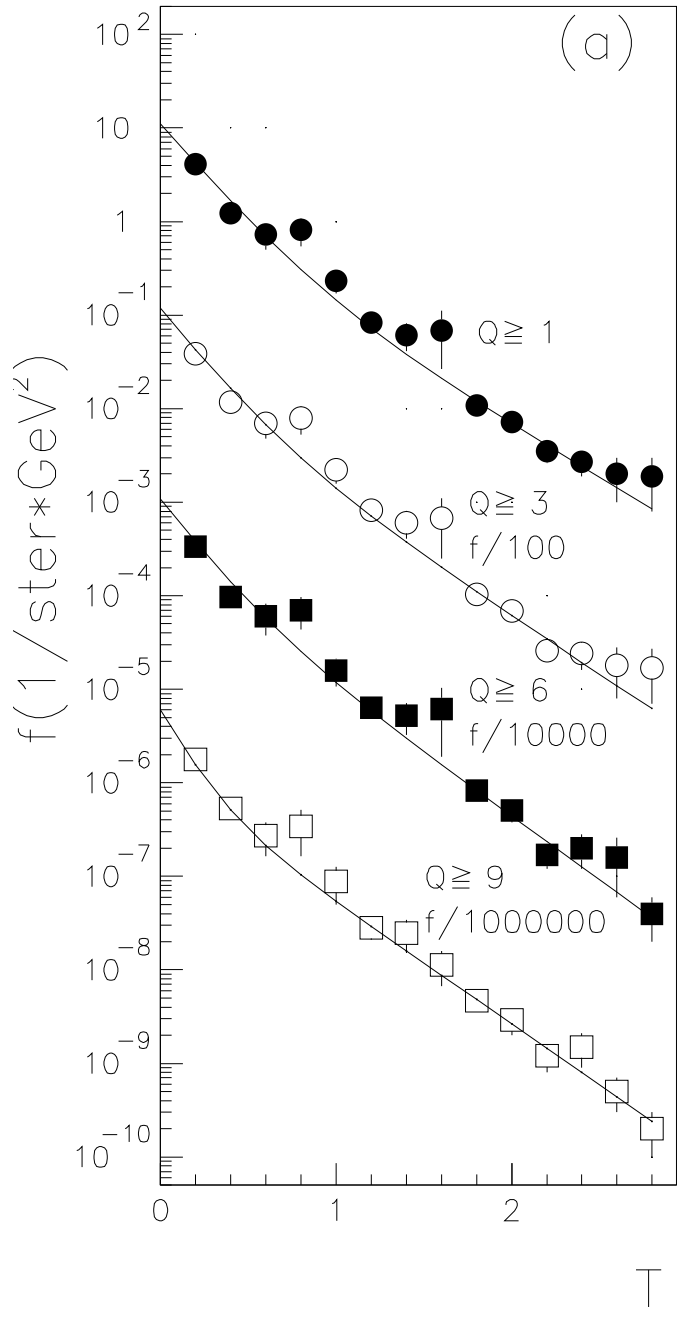
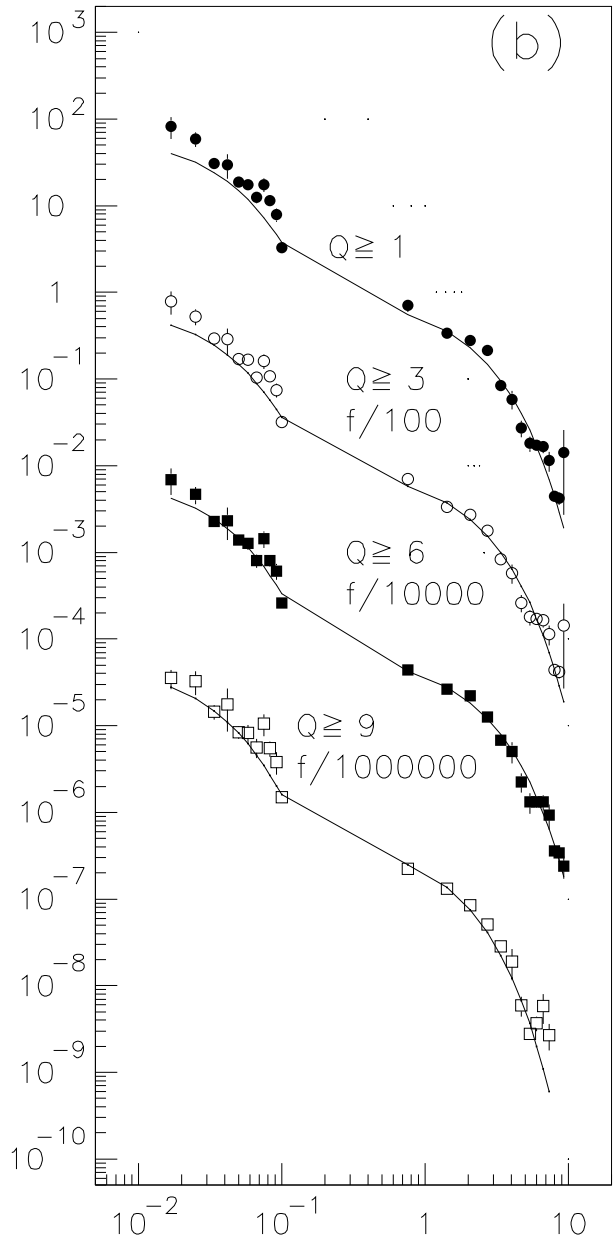
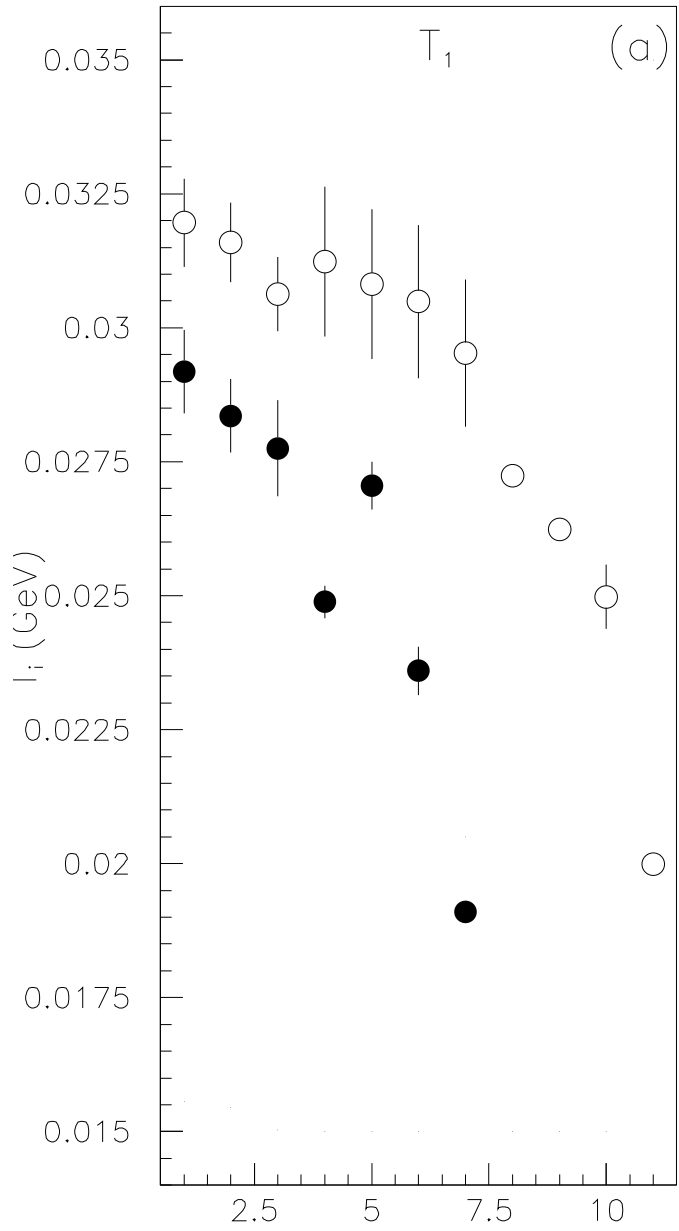
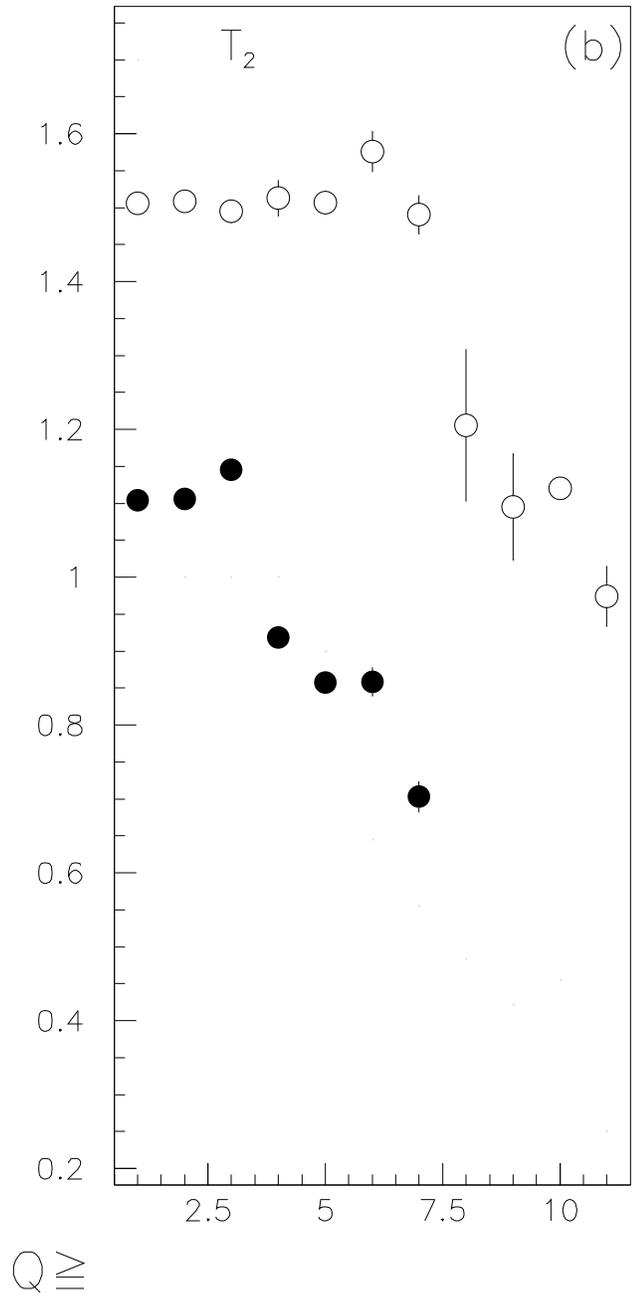
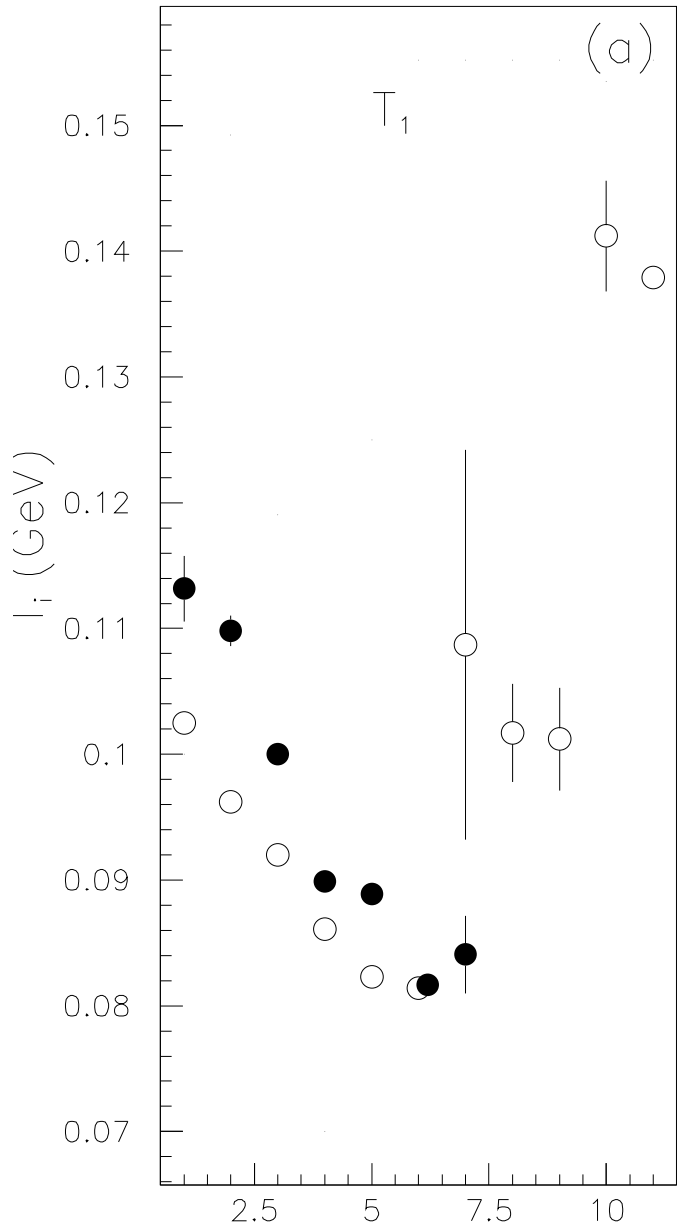
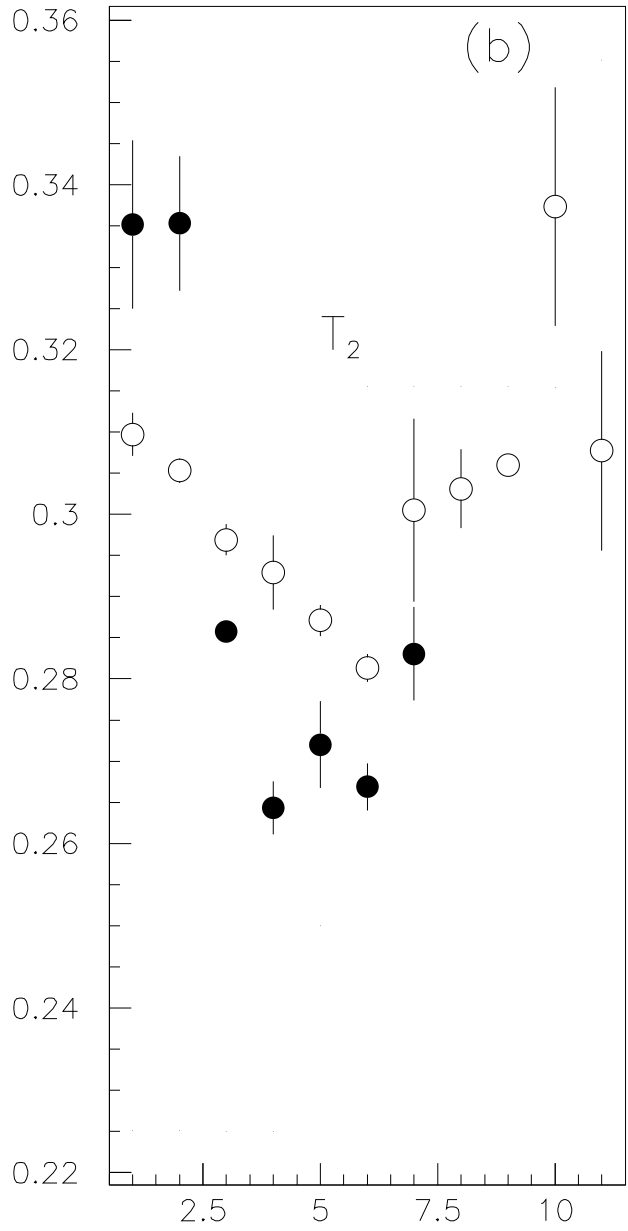
\end{document}